# Stock mechanics: a general theory and method of energy conservation with applications on DJIA

Çağlar Tuncay
Department of Physics, Middle East Technical University
06531 Ankara, Turkey
caglart@metu.edu.tr

**Abstract**

A new method, based on the original theory of conservation of sum of kinetic and potential energy defined for prices is proposed and applied on Dow Jones Industrials Average (DJIA). The general trends averaged over months or years gave a roughly conserved total energy, with three different potential energies, i.e. positive definite quadratic, negative definite quadratic and linear potential energy for exponential rises (and falls), sinusoidal oscillations and parabolic trajectories, respectively. Corresponding expressions for force (impact) are also given.

## 1. Introduction

A wide diversity of literature emphasizes randomness in short-range behavior of market. Yet, short- and long-range behaviors may differ in many characteristic aspects. We claim in the present work that, there may exist some fundamental and deep lying structures, which are not displayed in short-range fluctuations. Similar to the difference in several properties of shell and core electrons in atom, and the difference between surface and bulk properties of crystal, and so forth, the pronounced deeper structures in time series may be quite different than those of short-range states. We propose that, some quantities as a function of value (price) $\chi(t)$ of the index (share) and its $n^{th}$-time derivatives with arbitrary n, may be conserved during market journeys (composed of fluctuations, oscillations, rises and falls, crashes, etc). We will focus in the present work on these functions with only two arguments, namely $\chi$, and v (=$d\chi/dt$), i.e. speed. We will search the forms, where the variables are separated, and furthermore the speed part is ½$v^2$. These functions may be visualized simply as algebraic mappings of $\chi$-v space onto real number space. The quantity of ½$v^2$ and $\chi$-part may be called arbitrarily. Or, by assuming unit mass for value (price), one may define and consider ½$v^2$ term as the "kinetic energy", and the rest as the "potential energy", and the total as (we prefer to call) "total energy", respectively. In the existing perspective, $\chi$, v, a=$d^2\chi/dt^2$, etc. may be deduced in terms of analytical functions of these conserved quantities. And finally, the whole formalism may lead to describe the time series, epoch by epoch, where the corresponding potential energy expression remains constant in form and in parameters. Yet, there may be some other conserved quantities which have more complicated, i.e., non-linear, terms in $\chi$ and $d\chi/dt$.

In summary, our message is fourfold: Any time series may involve conserved quantities. Some of these quantities may be named (and considered) as energy, which may be utilized to obtain analytical expressions for price ($\chi$), time rate of change of price, i.e. speed (v), and acceleration (a=$d^2\chi/dt^2$), etc., as a function of time. Thirdly, long-range behaviors may be described epoch by epoch, each characterized by different potential energies involving different forms and parameters. And, the last, all these long-range properties may be quite different than those obtainable via short-range fluctuations.

*Kinetic, potential, and total energy*

One may define kinetic energy for prices $\chi(t)$ (taking mass equal to unity) as

$$K = \tfrac{1}{2}v^2 \quad , \tag{1}$$

where v is the usual speed i.e., $v(t)=d\chi(t)/dt$. Furthermore, a potential energy U in terms of difference in prices ($\chi(t) - \chi_{av}$) may be defined for any time series. In general, U is assumed to involve terms, which may be written as a sum like

$$U = h_0 + h_1(\chi - \chi_{av}) + h_2(\chi - \chi_{av})^2 + \Gamma(3) \quad , \tag{2}$$

where $\chi_{av}$ is some time average of prices, which defines the zero-potential-level if $h_0=0$, and $\Gamma(3)$ stands for the sum of third- and higher-order terms. The coefficients $h_n$ are proportional to price derivatives $d^nU/d\chi^n$ evaluated at ($\chi=\chi_{av}$) for n=0, 1, 2, and $h_1=h_2=0$, $|h_0/\Gamma(3)|<<1$ expresses a constant potential energy. Without any loss of generality, $\chi_{av}$ and $h_0$ in Eq. (2) is set to zero for any epoch.

We may write the total energy $E_T$ as

$$U + K + E_O = E_T \quad , \tag{3}$$

where $E_O$ represents any other form of energy (like Gibb's energy, entropy) may be present in the market, else then U and K. In this work we will confine ourselves to study only these epochs, for which $E_O \cong 0$, $\Gamma'(3) \cong 0$ ($E_O/E_T<<1$, $|\Gamma(3)/h_1\chi|<<1$, and $|\Gamma(3)/h_2\chi^2|<<1$) and assume conservation of the sum of potential and kinetic energies, as long as friction forces, damping etc. are negligible. The dimensional unit of energy terms (K, U, and E) may be taken as (local currency unit/time)$^2$ for shares, e.g., (¢/day)$^2$ or ($/day)$^2$ in USA. For indices local currency unit (lcu) may be kept in the units or it may be substituted by "value". In potential energy expression, $h_1$ and $h_2$ terms will have the unit (lcu/time$^2$), and the unit of (time$^{-2}$), i.e., frequency squared, respectively.

*Energy conservation and equations of motion*

Assuming conservation of total energy, time differentiation of Eq. (3) yields

$$h_1 + 2h_2\chi + dv/dt = 0 \quad , \tag{4}$$

where the common term v is dropped.

One has, for $|h_2/h_1|\ll 1$ and $h_2<0$ in Eq. (4) the familiar equation of motion for rises and falls in constant acceleration field, e.g. gravitation as in classical particle mechanics

$$\chi(t) = \chi_0 + v_0 t + \tfrac{1}{2} h_1 t^2 \quad, \tag{5}$$

where $\chi_0$, and $v_0$ designate initial price and speed, respectively. Time runs over exchange process days (t) and may be set to zero for the beginning of any epoch. For $|h_1/h_2|\ll 1$, and $0<h_2$ Eq. (4) yields oscillations;

$$\chi(t) = \chi_0 + v_{av} t + A\sin(wt + \Phi) \quad, \tag{6}$$

where A, w, and $\Phi$ is the usual amplitude, angular frequency (here, $(2h_2)^{\frac{1}{2}}$), and phase, respectively. Such oscillations are observed in general to have some medium range time periods, and fade away after a few (two, or three) full periods, where the sign and magnitude of $v_{av}$ indicates the up, down or horizontal character of oscillatory trends. For $|h_1/h_2|\ll 1$, and $h_2<0$ in Eq. (4), we have exponential growths,

$$\chi(t) = \chi_0 \exp(\beta t) + \chi'_0 \exp(-\beta t) \quad, \tag{7}$$

with $\beta^2=-2h_2$, and after a short transition interval the decaying part can be neglected.

It may be underlined that, U=0 corresponds to a zero impact stituation in market under the assumption of conservation of total energy, i.e., supply (offer) and demand meet and price is stable, since $F=a= -\partial U/\partial \chi$. The case $h_2=0$, and $|\Gamma(3)/h_1\chi|\ll 1$ describes a linear potential, (as gravity on the earth surface), where $h_1$ becomes the gravitational (anti-gravitational) constant with $0<h_1$ ($0>h_1$). For $h_1=0$, and $|\Gamma(3)/h_2\chi^2|\ll 1$ we have a quadratic potential energy, which may result in an exponential growth or oscillations for a spring-mass potential energy with a spring constant (Hooke constant) equal to $2h_2$.

The parameters of Eq. (2) i.e., $h_n$, may differ from one epoch to the other for any share and also from one share to another in any epoch. Whenever two nonnegligible terms exist in Eqs. (2) and (3), these cases can be treated by perturbational methods. Furthermore, we may give an accurate definition for epoch in energy language as, an epoch in a time series is the period of time when the (form of and the parameters involved in) potential energy remains the same. So, as long as the present form of U does not change, the present epoch survives, and vice versa. Furthermore, some additive potential energy terms may exist in the full expression, in order to involve the relatively small deviations (decorations) in price. Such cases, may be indicated as time terms within epochs.

It is worth to note that a similar (but not the same) approach has been followed by Ide and Sornette[1, 2], where the time derivative of speed was written in terms of a linear combination of $\chi^n$ and $v^m$, with arbitrary powers (See Eqs. (11)-(13), (32), (33), in [1], and Eqs. (1)-(3) in [2].) There, no assumption about the conservation of the sum of kinetic and potential energies was made. They have obtained oscillations and up (down) trends by neglecting the speed and restoring terms, respectively.

In the next section, DJIA is investigated in terms of epochs with different potential energies, where we ignore the day-to-day variations and claim only that long-term and medium-term trends of the DJIA conserve energy. Last paragraph is devoted to conclusion.

## 2. The index of DJIA

DJIA is selected here, because its time domain is the longest and volume is largest of all the world indices. And the whole history of DJIA, (Fig. 1.) from 01.10.1928 up to the present, can be decomposed into five main epochs, (utilizing daily real data in [3]) each with a different length of duration, as:

      1:      First 3,560 days (around Great Depression) till about 01.01.1943 ($\chi$=120.25). Transition and squezing the price are the main characteristics of this epoch.

      2:      From 01.01.1943 ($\chi$=120.25) till nearly 13.02.1964 (the $8854^{th}$ day, $\chi$=794.42). Duration is about 5,294 days. Exponential growth with the exponent (=$\beta$) nearly equal to 0.0003 $day^{-1}$ is the main characteristic of this epoch.

      3:      From 13.02.1964 till 01.01.1982 ($13340^{th}$ day $\chi$=1000) Duration is about 4,488 days. Oscillations with angular frequency (=w) equal to 0.0084 $day^{-1}$ between 750 and 1,000 levels is the main characteristic.

      4:      From 01.01.1982 till 2000 crash ($18000^{th}$ day). Duration is about 4,660 days. This epoch consists of two long terms (partial epochs), each displaying exponential growth with different exponents of 0.0005 ($day^{-1}$) between 01.01.1982 and 08.12.1994, and 0.0008 ($day^{-1}$) afterwards till 2000 crash, all respectively.

      5:      Current epoch. Rough constancy about the index level of 10,000 is the main characteristics of this epoch.

It is worth to note that, the pronounced epochs do not seperate from each other sharply, but some transition terms occur in general between them.

**First epoch and the fundamental line of DJIA:** Referring to Fig. 1., which is obtained utilizing daily real data in [3], the first epoch is the transitional period and lasted about 3,565 days around Great Depression. The main characteristics of this epoch is triangular squeezing of the prices with the climax meeting the beginning of the Sept.1929 crash at a value of 381.17, which ended after about 700 days at a value of 41.22. Then a sharp rise occured till the value of about 200. Afterwards some decrease came reaching its deepest value at 92.92. These two successive local minima, i.e., 41.22 and 92.92 are very important figures in the history of DJIA since, the line passing through them defines the lower rising edge of the pronounced triangle and moreover the rightward extension of this line matches with the minima of prices for many decades, till the value increases to 1,000. (See, the longest straight line in Fig. 1.) The exponential increase, corresponding to the long straight line in Fig.1 is in fact worth to be considered as the fundamental line of DJIA, because of several reasons. First of all, this line can be taken as the $0^{th}$-order approximation for the index, since the corresponding time series is observed to evolve about (below and above of) it. Secondly the fundamental line, has played the role of supporting line till the date of 16.June.1989 ($10,172^{th}$ day). The day after, i.e. at 17.June.1989 the index has broken down from it (after many previous hits) and later has never crossed it. Starting in Apr.1999 DJIA has approached it from below many times, (at index values between 11,000 and 11,700) till the end of the same year, but has ceased to override it, and the last log-periodic crash has followed. (For detailed studies of this crash, in a power law formalism, see [4-23].) The log-linear slope of increase of the fundamental line is calculated to be $\beta$=ln(92.92/41.22)/(3390-946)=0.000333$day^{-1}$. The $0^{th}$-order potential energy and the corresponding equation of motion, utilizing the conservation of total energy, for DJIA may be written as

$$U^{zeroth\text{-}order} = h_2\chi^2 = -0.5\beta^2 \exp(2\beta t) + C \quad , \tag{8}$$

and

$$\chi(t)^{\text{zeroth-order}} = \chi_0 \exp(0.000333\ t)\ ,\tag{9}$$

respectively. In Eq. (8) the constant C takes care of the initial potential energy and it can be absorbed within U, and in Eq. (9), $t_0=0$ is set, as can be done anew for each epoch. Note that, we have the price velocity and acceleration as $v=\beta\chi$, and $a=\beta^2\chi$, respectively, and both also increase exponentially in time. The $0^{\text{th}}$-order force (and so, the acceleration) expression may also be written as $F^{\text{zeroth-order}} = -\partial U/\partial\chi = (0.000333)^2 \chi_0 \exp(0.000333\ t)$. Yet, it is obvious that the pronounced approximation can not be considered as a very successful one for the whole history of DJIA (Fig. 1.).

As a better approximation, one may try Detrended Fluctuation Analysis (DFA) onto index values. DFA is a common method in literature, widely used to eliminate effects of short-range fluctuations in computations. In DFA, the time series is considered segment by segment, and averages are computed by integrating prices over each segment and then dividing the result by the length of time duration. In the present application of DFA, it is utilized with moving averages, where a linear-DFA with running average is performed. DFA line could barely be distinguished from the real data and is thus not shown in Fig.1. The present DFA curve is definitely not unique, i.e. infinitely many similars may be obtained on different criteria, such as accuracy wanted in results, limiting lengths of each segment to be less than a desired value, etc. (For the details and several applications of DFA, see [24-34].)

**Second epoch and psychological resistances:** It is also displayed in Fig. 1. that, exponential rise in $\chi(t)$ survives till it reaches the value of 1,000, i.e., upto a "psychological resistance". Many people believe that such round figures in share prices and index values play the role of resistance, and maybe their belief is what makes these figures resistive in reality. In any case, it can be observed that there already exist three such "round resistance" in DJIA; namely, 100, 1,000 and 10,000 as shown (by horizontal lines for 1,000 and 10,000 resistances) in Fig. 1., around which price displays the habit of spending some thousand days. The second epoch, between the initial transition period and the 1,000 resistance level displays exponential growth as described in Eq. (9) with the potential energy given in Eq. (8), since the fundamental line passes through the minima of the second epoch.

**Third epoch:** The term taking place between the 1,000 resistance horizontal line and the horizontal line at the level of 750, may be considered as being composed of six sinusoidal oscillations with a period of T=750 day (Fig. 2.). This period corresponds to $h_2 = 3.5 \times 10^{-5}$ (rad/day)$^2$ of the Eq. (2), and with A from Eq.(6) taken as 125 lcu, its full expression may be written as

$$\chi(t)^{\text{epoch 3}} = 875 + 125\sin((2\pi/750)t - \pi/2)\ ,\tag{10}$$

where $v_{av}=0$, (Eq. (6)), and $t_0$ is taken at the date of 11.Apr.1963. It can be shown that, Eq. (10) is the resulting equation of motion with a positive $h_2$ in the potential energy (Eq. (2)) of

$$U^{\text{epoch 3}} = h_2\ (\chi(t)^{\text{epoch 3}})^2\ ,\tag{11}$$

where the constant term corresponding to $\chi_{av}$ = 875 lcu is absorbed in U. Then, total energy $E_T$ can be calculated as $E_T = \frac{1}{2}v^2 + \frac{1}{2}(2h_2)\chi^2 = 2(\pi A/T)^2 = \pi^2/18 = 0.55$ (lcu/day)$^2$, which is conserved throughout the whole epoch.

Oscillatory epochs end up with an abrupt change, in general. The third epoch, surviving about 5,000 trading days (nearly 20 calender years) did not alter the "rule", and terminated with an abrupt rise. And so, the third epoch is closed and the gate to the fourth one is opened.

**Fourth epoch:** We meet exponential rise again, with a steeper slope this time (arrow 2, in Fig. 1.), with 0.0005 lcu/day (=ln(3674.63/776.92)/(16,603-13,496). The equation of motion is

$$\chi(t)^{epoch\ 4} = \chi_0^{epoch\ 4} \exp(0.0005\ t)\ , \qquad (12)$$

where $\chi_0^{epoch\ 4}$ may be taken as 776.92 at 12.Aug.1982 as the 13,494$^{th}$ day (counting from the date 01.10.1928 on), which must be set now as the first day for the fourth epoch. Corresponding potential energy becomes as

$$U^{epoch\ 4} = -1.25 \times 10^{-7} \exp(0.001\ t)\ , \qquad (13)$$

where the constant term is dropped in usual manner. In fact, this epoch may be considered (in more details) as composed of two partial-epochs, with two different slopes, as discussed right after the beginning of this section.

**The final (current) epoch** is, naturally, the most interesting one of all considered here, since it is not completed yet. Will it oscillate many times again as it did in the third epoch, or will it follow a different evolution nowadays? Certainly, nobody can answer this and similar questions with certainty. Yet, several scenarii may be proposed relying upon several criteria.

Firstly, as a personal feeling it may be phrased that, the present epoch will not be of sinusoidal oscillatory type. Because, the maximum of the second rise beginning in Apr2003 is below the previous one as can be seen in Fig. 3. Moreover, the second maximum is much lower than the first one in S&P500 and NASDAQ too, which had displayed sinusoidal oscillations in the same time period with the third epoch of DJIA. Resting on this perspective, we suggest a gravitational potential energy of the form $U=h_1\chi$ to describe the current stituation. The general equation of motion deduced out of the given potential form is the Eq. (5). Utilizing the real data [6], we obtain (See Fig. 1 in [34])

$$\chi(t)^{epoch\ 5} = 6415.3 + 10.17t - 0.00606\ t^2\ ,$$

$$\chi(t)^{epoch\ 5} = 7692.1 + 8.43\ t - 0.00606\ t^2\ , \qquad (14)$$

for the first and second portions (with respect to the local minimum at the beginning of 2003), respectively. Notice that the gravitational acceleration of $h_1/2=-0.00606$ lcu/day$^2$ remains the same for both of the parts, i.e., full time domain (8 calender years) of the current epoch.

Is this an accidental result? We obtained the similar constancy of the acceleration for S&P500, and NASDAQ (See Figs. 1-3 in [34]). So, the stituation may not be considered as accidental result, since it occurs in three different indices for sufficiently long time domains.

Within the existing perspective, one may imagine a price particle shot upward at a height of 6415.3 lcu with an initial speed of (rounding the speed figures to two digits) 10.17 lcu/day in a downward gravitational field of $|h_1|$. It reaches the first maximum, which can be calculated utilizing $\chi_{max} - \chi_0 = v_0^2/(2|h_1|)$, the very familiar formula from the physics of particle kinematics. Afterwards, it starts to fall freely back to the initial level, and strikes with the (initial) speed of 10.17 lcu/day. This collision may not be considered as an elastic one, since the departure (bouncing back) speed of 8.43 lcu/day is less than the original one, accounting to 30% loss in kinetic (and total) energy. It already (at the date of March2005) arrived the second maximum, which can again be calculated utilizing the equation of $\chi_{max} - \chi_0 = v_0^2/(2|h_1|)$, after substituting the new speed for $v_0$. A new free fall back to 8,000's at least, may be expected within the coming 500 days... Similar interpretations and predictions about S&P500 and NASDAQ may easily be developed as done in [34]. S&P500 and NASDAQ are expected to recede back in 500 days, to 800's and 1,400's, respectively. As the second scenario, it may be stated that, if a price squeezing occurs below the line passing through the local maxima after 2000, DJIA may very well display a new transitional period in future of about some years, except for some short time periods. As the third scenario, DJIA may enter into a new sinusoidal oscillatory epoch about the 10,000 index level, for some years again, as it had sailed horizontally below the 1,000 level; provided it rises up sharply towards 12,000 within a few months. In fact, this choice can be considered as a more consistent one with the previous epochs, (with respect to, constant gravitational acceleration case, especially) since, oscillations have quadratic potential energy, and quadratic potential energy is the dominant form in DJIA. With respect to any of the three scenarii discussed above, DJIA (and so, S&P500 and NASDAQ) may be expected to fall in the coming months.

*Energy conservation in DJIA*

In the previous paragraphs, the given potential energies for each epoch were assumed without any proof, except that the resulting equations of motion suit well to describe the characteristic behavior of each epoch. By utilizing DFA with moving averages, we calculated fluctuation detrended price. And utilizing this price we calculated the kinetic and potential energies and their sum for each epochs. Many of the aspects of each DJIA epoch (except the first epoch, where not any potential energy expression is attained due to transition) may be overviewed in energy language.

**Second epoch:** Due to exponential growth, as the main characteristics of this epoch, the potential energy is quadratic with a negative factor, which is equal to half of the minus the square of the exponential. And the total energy becomes zero. Utilizing DFA segments centered at each t (day), we obtained that total energy is fluctuating in very small amounts about zero. As we approach to the next epoch the pronounced fluctuation increases. This term may be considered as the transition term between the second and the third epochs. The DFA segments utilized, have 100 days of half length, which are almost uniformly distributed on t.

**Third epoch:** Almost sinusoidal oscillations, as given in detail in Fig. 2., involve a quadratic potential energy with a positive multiplier being equal to half of the square of the angular frequency associated. So, the energy is nonzero and equal to 0.55 ($lcu^2/day^2$), as described in the paragraph just following Eq. (11) above.

**Fourth epoch:** Moving averages, (which could barely be distinguished from the real data and are thus not shown in Fig.3.) are obtained utilizing DFA with segments of half length equal to 500 days maximally, where the exponent is taken as equal to 0.0005 ($day^{-1}$) for the whole epoch. (Yet, the time term after 01.01.1966 has a steeper increase, which is indicated by a thinner line of arrow 2 in Fig. 1. As a result, total energy calculated with the pronounced exponential, ceases to be valid in time term after 01.01.1966.

**Fifth epoch:** The gravitational potential energy involved in the current epoch is equal to $0.01212\chi$, where $\chi$ is measured with respect to the 7,500 level of index, since the epoch starts to take place there. Therefore, the total energy, being equal to gravitational potential energy at maximum price height, is about 36 ($lcu^2/day^2$). The outcoming total energy in our computations is found to be fluctuating in small amounts about the level of 36 ($lcu^2/day^2$), proving roughly the conservation.

In summary, we have the total energy as roughly conserved in each epoch, where the second and the fourth epochs involve zero and the third and the current epochs, non-zero and positive total energies. The moving average curve utilized in the present DFA analysis is not unique and infinitely many of similars to it may be utilized in the original formalism. So, the proposed method is general and it is not constrained to any fittings, provided the characteristic parameters are given. Yet, if the pronounced parameters are utilized with wrong values, the correponding invalidity may be easily and directly detectable by the method. In this manner, one may even scan all the possible values for each parameter in blindfold manner, and investigating the outcoming results, one may pick up the correct values.

**Acknowledgement**

The author is thankful to Dietrich Stauffer for his friendly, discussions and corrections, and supplying some references.

**Figure captions**

**Fig. 1.** DJIA, with logarithmic price axis. Five fundamental epochs composing the history can be selected. The first "transitional" epoch is characterized by its triangular squeezing the price about the first "psychological resistance" of the value of 100. In the second and the fourth epochs log-linear (i.e., exponential) ascending of the price can be distinguished (arrows 1 and 2). In the third epoch, just below the second "psychological resistance" of the value of 1,000 oscillations are dominant. In the fifth, and the final epoch, the evolution of formation (about the third "psychological resistance" of 10,000 index) is not completed yet.

**Fig. 2.** The oscillatory third epoch of DJIA, taking place just below 1,000 and above 750 index values. Sinusoidal fit may not be considered as perfect, yet as the period of any oscillation elongates, the corresponding amplitude increases, and $E_T = 2(\pi A/T)^2$ remains the same. Oscillations terminate with an abrupt rise.

**Fig. 3.** DJIA about the Apr-Sep2000 climax is (ignoring short-term fluctuations) described by a rise and a fall of the price in a gravity $h_1/2 = -0.00303$ (lcu/day$^2$). The initial (shooting) speed at the beginning of 1995 is $v_{01} = 10.17$ (lcu/day). The price fall down after the maximum height and inelastically bounces back with $v_{02} = 8.43$ (lcu/day) in the same gravity, and rises up to 11,000's in accordance with the expression $\chi - \chi_0 = v_0^2/(2|h_1|)$. A recession, back to 8,000's and below is expected within the coming 500 days.

**FIGURES**

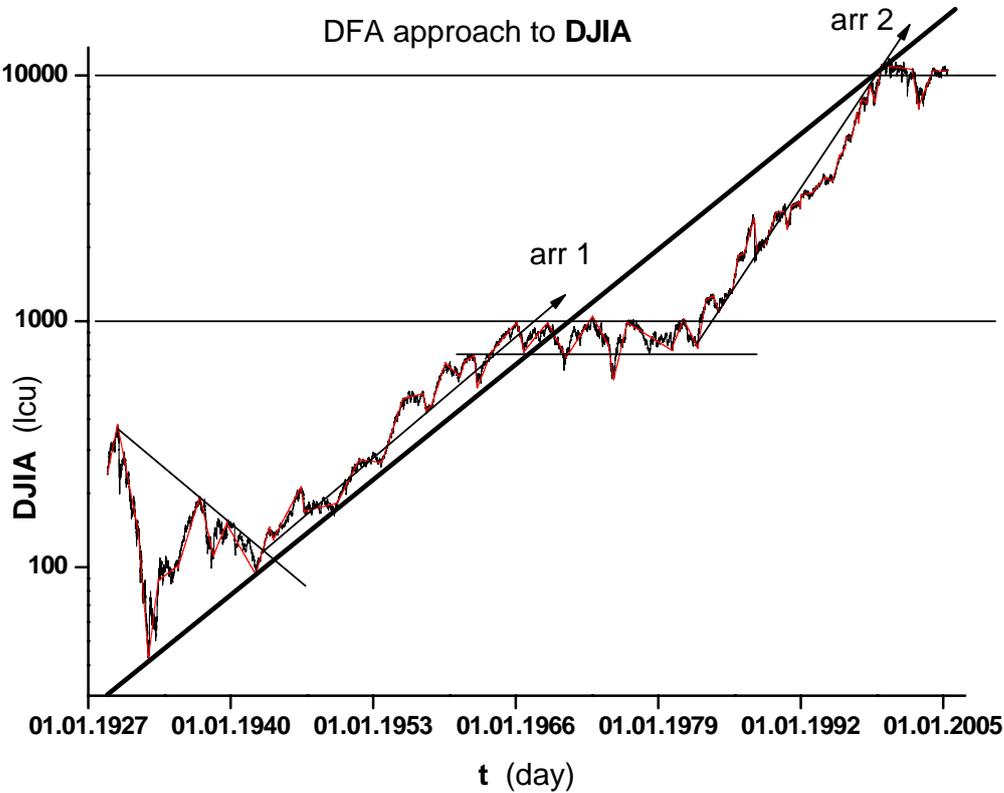

**Fig. 1.**

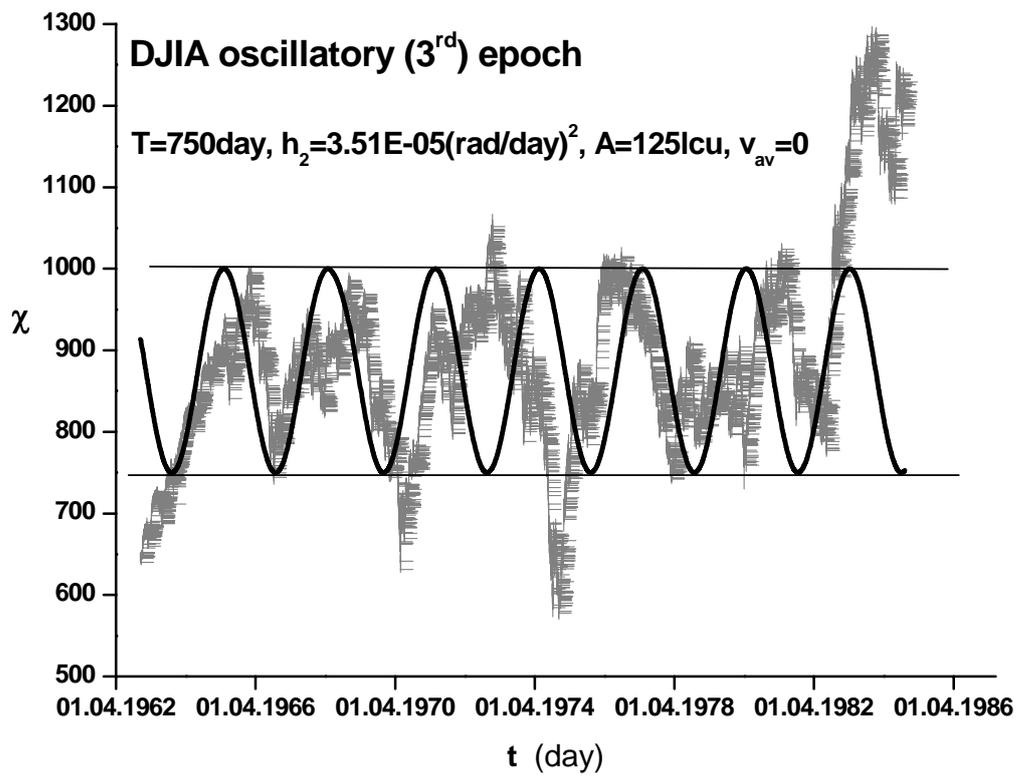

**Fig. 2.**

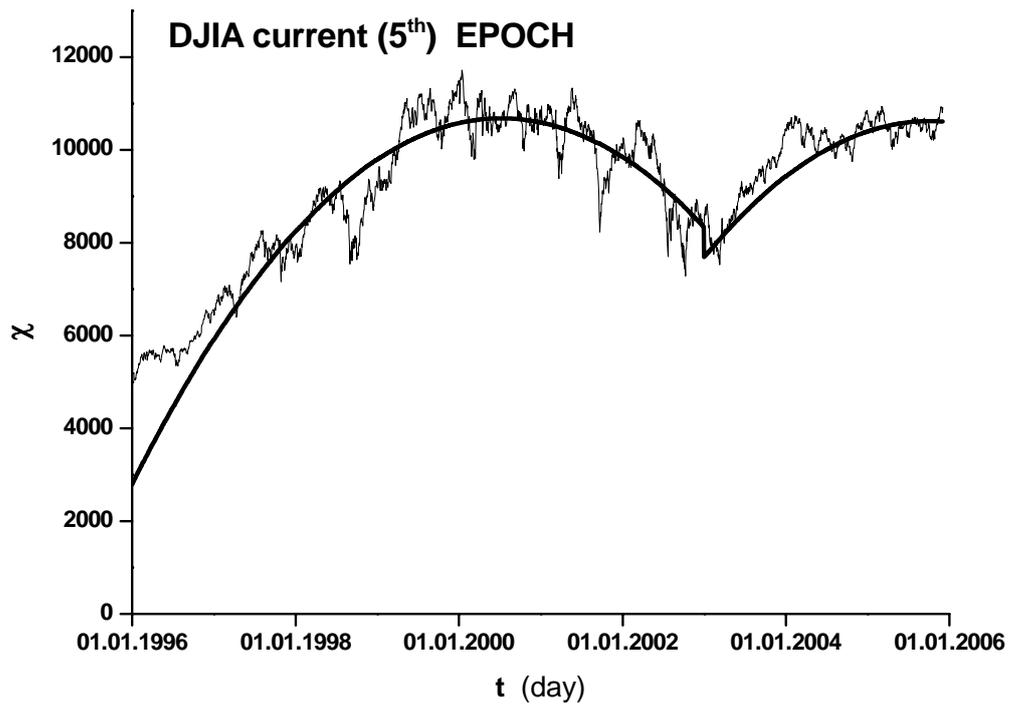

Fig. 3.